\documentclass{aa}
\usepackage{graphicx}
\usepackage{amssymb}
\begin{document}
\title{High precision determination of the atmospheric parameters and 
abundances of the COROT main targets}
\author{M.  Gillon \inst{1} \and   P.  Magain \inst{1}}
\offprints{M.  Gillon\\email : gillon@astro.ulg.ac.be}
\institute{Institut d'Astrophysique et de G\'eophysique,  Universit\'e 
de  Li\`ege, All\'ee du 6  Ao\^ut, 17, Bat.  B5C, Li\`ege 1,   Belgium }
\date{}
\abstract
{One of the main goals of the  COROT   mission is to get   precise  
photometric observations of selected bright stars in order to allow the modelling 
of their interior  through  asteroseismology.  However, in order to
interpret the asteroseismological data,   the effective temperature,
surface gravity,  and chemical composition of the stars must be known with
sufficient accuracy.}
{To carry out this task, we have developed a spectroscopic method called   
APASS (Atmospheric   Parameters  and  Abundances from Synthetic Spectra) 
which allows precise analysis of stars with a  moderate to 
high  rotational velocity, which is the case for most primary COROT targets.}
{Our method is based on synthetic spectra in which individual lines are replaced
by \emph{analysis units} (isolated lines or line blends, depending on
the crowding of the spectral region and on the rotational broadening). 
It works differentially with respect to the Sun and allows the atmospheric 
parameters and chemical abundances to be determined by considering analysis
units with different sensitivities to these various parameters.}
{Using  high  signal-to-noise  spectra and the APASS method,    we    
determined  the atmospheric parameters and  chemical abundances  of  13 primary
COROT targets.  Our results agree well with  those  obtained   by  Bruntt 
using his  software VWA and with those obtained  with  the   software  
TEMPLOGG. However, in both cases,  our error bars are significantly smaller than those
of other methods. Our 
effective temperatures are also in excellent  agreement  with  those 
obtained with the IR photometry method. For five stars with  relatively 
low rotational  velocity,  we also  performed  an  analysis with a classical 
equivalent--width  method  to  test  agreement with APASS  results. 
We  show that equivalent--width measurements by Gaussian or    Voigt  
profile--fitting   are  sensitive  to   the  rotational  broadening,  
leading   to systematic  errors whenever the projected rotation velocity
is non--negligible.  The APASS method appears superior in all cases and
should thus be preferred.}
{}

\keywords{stars: abundances -- stars: atmospheres -- stars: fundamental 
parameters}
\authorrunning{M. Gillon \& P. Magain}
\titlerunning{Atmospheric parameters of COROT main targets}
\maketitle
%
%
\section{Introduction}
COROT (COnvection, ROtation and planetary Transits)  will   carry  out 
several  projects  in asteroseismology (\cite{Baglin}). It will also be 
the first satellite launched  with the  aim of    detecting exoplanets 
by  the transit method (\cite{Rouan01}). The instrument features  a 
27 cm telescope, and the  asteroseismology    program    will  use  two 
2048$\times$2048 CCD  cameras, leading to  a  field  of  view  of  3.5
deg$^{2}$. During the 2.5 years  of  the  mission,  5  fields  of  the
Galactic  plane  will be  observed  continuously, each  one during 150
days. Among these fields, a  few  bright  stars have been selected  as 
candidate primary  targets of the  mission,  and  getting   a   precise 
photometric monitoring of these stars to allow the modelling  of their 
interior is the main goal of the  asteroseismological  part  of  COROT 
(see Table 1 for the preliminary list of these primary targets). 

To   allow    thorough   modelling   of   the   stellar     interior, 
asteroseismological data are not enough, so additional constraints must be 
provided by precisely determining the atmospheric parameters and
abundances. To accomplish this task, we set up a spectroscopic  analysis 
method, APASS (Atmospheric  Parameters  and  Abundances  from Synthetic 
Spectra), that aims at the highest possible 
precision. With this method, we  analysed the  majority of  the 
COROT  main  targets, including some stars later  rejected  from the 
preliminary main  target  list.

In Sect. 2, we present the spectroscopic observations used in this work, and in
Sect. 3,  the  method APASS is described. In Sect. 4, we   present the
results  and  compare them to   previous  studies.  We 
give our conclusions in Sect. 5.

\begin{table}
\centering
\caption{Preliminary list of COROT main targets  from   \cite{Baglin1}.
The $v$ sin$i$ values were provided by C. Catala (private  communication). 
HD 52265 has a giant planet detected by spectroscopy (\cite{Butler}).}
\begin{tabular}{cccc}
\hline\hline Star & Spectral type & $v$ sin$i$ (km s$^{-1}$) &\\
\hline
HD 52265 & GO III-IV & 5.2 &  \\
HD 49933 & F2 V & 10.9 & \\
HD 49434 & F1 V & 89.6 & \\
HD 43587 & F9 V & 5.9 & \\
HD 170580 & B2 V & 16.0 & $\beta$ Cep\\
HD 171834 & F3 V & 72.1 &\\
HD 177552 & F1 V & 41.2 &\\
HD 181555 & A5 V & $>$ 200 & $\delta$ Scuti\\
HD 180642 & B1.5 II-III &  & $\beta$ Cep\\
HD 45067 & F8 V & 8.0 &\\
HD 46558 & F0 V & 63.1 &\\
HD 43318 & F6 V & 7.5 &\\
HD 175726 & G5 V & 13.5 &\\
HD 171234 & A3 V & 162.8 &  $\delta$ Scuti\\
\hline
\label{tab:ere}
\end{tabular}
\end{table}

%
%
%
\section{Spectroscopic observations}

We  extracted a spectrum of each of the COROT   main    targets 
analysed  here  from  the COROT ground-based asteroseismology database,  
Gaudi\footnote{http://sdc.laeff.esa.es/gaudi} (\cite{Solano}).  All but
one  were obtained with the ELODIE  spectrograph attached to the 1.93 m 
telescope at Observatoire   de    Haute-Provence (OHP).  ELODIE  is   a  
fiber-fed cross-dispersed Echelle  spectrograph that  provides  a complete  
spectral coverage  of  the   3800-6800  $\AA$  region, with a resolving 
power R = 45000 (\cite{Baranne}). The spectrum of HD 52265 was 
obtained with the CORALIE spectrograph  attached to  the 1.20  m  Swiss 
telescope at La  Silla  (Chile), providing  a   resolving   power   R =   
50000 (\cite{Queloz}). For one star, HD 49933, we also made an 
analysis  with a co-addition of $\sim 100$  spectra obtained using  the  
ultra--high resolution (R = 115000) HARPS spectrograph attached to  the  
3.60  m ESO telescope at  La Silla (\cite{Mayor}).  Table~2  gives  the  
observation log for the spectra used in this analysis.

\begin{table}[t!]
\centering
\caption{Observation log for the  spectra  of  the stars 
we analysed. The signal-to-noise  ratio
($S/N$) in   the last  column    is  calculated    around    6500   $\AA$. 
$co$-$ad$: co-addition of $\sim 100$ spectra taken the same night. }
\begin{tabular}{ccccc}
\hline\hline Star & Instrument & Date  & UT start&  $S/N$\\
\hline
HD 43587&ELODIE&14-Jan-98&22:32&250\\
HD 57006&ELODIE&10-Dec-00&00:41&250\\
HD 45067&ELODIE&15-Jan-98&22:20&260\\
HD 43318&ELODIE &19-Jan-98&22:48&120\\
HD 49933&ELODIE&21-Jan-98&22:17&210\\
HD 49933&HARPS&19-Jan-04& $co$-$ad$ &750\\
HD 55057&ELODIE&15-Jan-98&23:10&270\\
HD 184663&ELODIE&18-Jun-00&01:30&170\\
HD 49434&ELODIE&17-Jan-98&22:33 &160\\
HD 171834&ELODIE&05-Sep-98&19:00&300\\
HD 177552&ELODIE&14-Jul-98&22:46&260\\
HD 175726&ELODIE&10-Jul-01&21:23&140\\
HD 46558&ELODIE&21-Dec-99&01:08&140\\
HD 52265&CORALIE&24-Nov-01&08:03&140\\
\hline
\label{tab:rer}
\end{tabular}
\end{table}

\section{Description of the method}

Using the assumptions of Local  Thermodynamic  Equilibrium  (LTE)  and
plane-parallel,   homogeneous  atmosphere,   APASS  is  a differential 
 method based  on  the synthesis of selected spectral regions with 
Sun-calibrated line data and comparison with  a  stellar
spectrum in order  to  determine  the  projected  rotational  velocity 
($v$ sin$i$),  the effective temperature  $T_{\rm eff}$, the   surface 
gravity log $g$, the microturbulence velocity $\epsilon_{\rm turb}$, as 
well as the atmospheric abundances. 

\subsection{Analysis Units (AnU)}

We  developed APASS in order to be  able  to  precisely  
analyse stellar spectra with a moderate to high projected rotational 
velocity (see Table~\ref{tab:ere}),  which  cannot  be analysed with a 
classical equivalent--width (EW) method. Such spectra are convolved with  
the rotation profile, and  most  lines  are  blended  with  other  ones, 
forming ``packs'' of lines. One of the main  characteristic  of APASS is 
that it considers  either   individual  lines  or  ``packs''  of  lines,
treating them as  analysis  units  (AnU). Those AnU are selected in  the  
first  step  of  the  analysis, by visual  inspection   of  the  stellar 
spectrum. All  the AnU that are dominated  by iron (iron-dominated AnU 
: Fe--AnU) are used in determining  the atmospheric parameters,
while the other AnU are used to determine the atmospheric abundances.

\subsection{Atmospheric models}

We used models interpolated  in  the  Vienna  model  atmosphere    grid 
(\cite{Nend1}, \cite{Nend2}; \cite{Heiter}). They were  computed with a  
modified version of  the  ATLAS9  code  (\cite{kurucz2})  in     which 
turbulent convection  theory  in  the   CGM formulation (\cite{Canuto}) 
is implemented.

\subsection{Atomic data}

APASS uses limited   spectral   regions  selected on the basis of  the 
following criteria :\begin{itemize}
\item scarcity of telluric lines,
\item absence or scarcity of molecular lines,
\item high diversity of atomic lines: different elements, various 
sensitivities  to  the atmospheric parameters,
\item presence of numerous pseudo-continuum windows to accurately
normalize the continuum in order to gain  precision.
\end{itemize}

The  atomic  lines  data  were  extracted  from  the  VALD     database 
(\cite{kupka1}; \cite{Piskunov1}; \cite{Rya01}) and the Kurucz   atomic
lines database (\cite{kurucz1}). For all the lines with  a  significant 
equivalent--width, we      made a new  determination  of the    log $gf$ 
from an  inverted   analysis    of the   Jungfraujoch  Solar      Atlas 
(\cite{Delbouille}), using a CGM solar model having ($T_{\rm eff}$, log
$g$, $\epsilon_{\rm turb}$) = (5778 K,   4.44 dex,   1.00 km s$^{-1}$) 
and  solar  abundances   from \cite{Grevesse}.  Whenever  possible,  we
used   damping    constants    calculated following Anstee  \&    O'Mara 
(\cite{Anstee1}),     which     have     proved   to be quite accurate 
(\cite{Anstee2}; \cite{Barklem}). 

\subsection{Parameter determination}

APASS is an iterative method. Starting  from  an  initial  model,   the 
method consists of  cycles  of   determination   of   the   parameters, 
repeated until  convergence  for all parameters. Each   cycle  includes 
adjusting  the continuum,  determining the level   of  
convolution  of the spectrum,  and determining  the  atmospheric  
parameters  $T_{\rm eff}$, log $g$, $\epsilon_{\rm turb}$,  and  [Fe/H]. 
The projected rotational  velocity $v$ sin$i$ is determined afterwards. 

\subsubsection{Initial model}

For each star analysed in this work, we   begin    with   an   initial 
atmospheric model computed  with  parameters  $T_{\rm eff}$,  log $g$, 
$\epsilon_{\rm turb}$ and a metallicity extracted from the  literature 
or obtained using Str\"omgren indices  from the Gaudi database and the 
sofware  TEMPLOGG (\cite{Rogers}). To  synthesize the spectrum of  the 
selected  regions,  we  also need to  set  initial  values  for  two 
parameters representing the convolution of the spectrum: 
\begin{itemize}
\item the projected rotational velocity $v$ sin$i$. As  initial 
value, we use the $v$ sin$i$ determined  by C. Catala (private communication), 
\item the  parameter $\Omega$ representing the combined effect of the 
instrumental profile and the stellar  macroturbulence.  We   use   an 
initial value  of  4   km s$^{-1}$ for $\Omega$ and assume a Gaussian 
shape for that convolution profile.
\end{itemize}

\subsubsection{Correction of the Doppler shift}

Before starting the first cycle of parameter determination, 
we use the synthetic spectrum computed  with  the  initial   model  to 
determine the Doppler shift of the stellar spectrum. For this purpose, 
we select a large number of AnU and  determine the ratio of their 
position in the stellar spectrum and in  the  synthetic spectrum. We  use  
the average ratio for correcting   the  spectra that are synthesized 
afterwards.

\subsubsection{Adjustment of the continuum}
 
We begin each cycle with a new determination of  the  continuum of the
observed  spectrum on the basis of a number of pseudo-continuum windows
selected from inspection of both the observed  and  synthetic spectra. 
The mean  flux in these windows is measured in  both the   theoretical 
and the observed spectra, and a table is  constructed that  contains  the 
ratios of the mean  fluxes. A spline  curve  is  then  fitted  through
these points, and   the  stellar  spectrum  is divided by  that  curve, 
providing   our  re-normalized  spectrum. This method allows us to 
take into account the weak lines present in
the pseudo--continuum windows as accurately as possible.

\subsubsection{Determination of the convolution parameter $\Omega$} 

\begin{figure}
\centering  \rotatebox{270}{
\includegraphics[width=6.5cm]{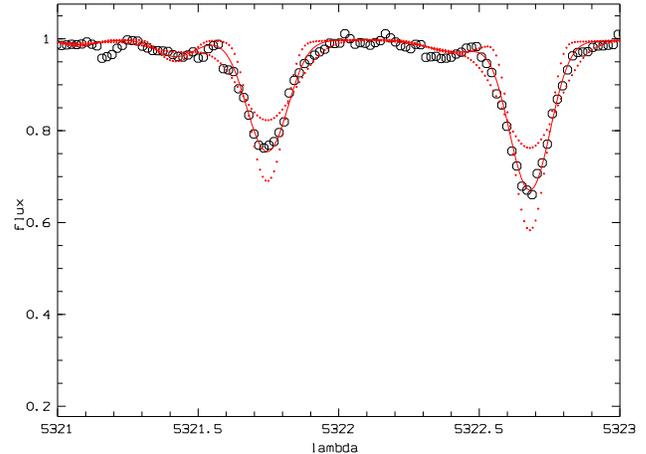}}
\caption{Portion of the stellar spectrum of HD 52265 (empty 
circles) showing two   Fe--AnU.   Overplotted:  synthetic  spectrum 
computed with $\Omega$ =  0.75    and   6.75  km  s$^{-1}$   (dotted lines)
 and with $\Omega$ = 3.75 km s$^{-1}$, the best--fit value 
(continuous line). \emph{lambda} = wavelength in $\AA$, \emph{flux} 
=  normalized stellar flux. }
\label{fig:lkp}
\end{figure}

The convolution  parameter $\Omega$ (macroturbulence +  instrumental
profile) is  determined for every  Fe--AnU   by   adjusting the  
synthetic   spectrum with an  $\chi^2$ minimization  method, and  the  
average overall Fe--AnU is the adopted value for  $\Omega$   (see  
Figure~\ref{fig:lkp}). During all  iteration cycles,  we  keep $v$ 
sin$i$ fixed at its initial value. Indeed, as these two
broadening parameters appear to be strongly anticorrelated in an 
$\chi^2$ minimization, allowing both of them to vary produces 
instabilities in the algorithm.

\subsubsection{Determination of the Fe abundance}

A value of the Fe abundance is deduced  from  each  Fe--AnU  in  the 
following way. First, the user defines the limits in  wavelength  of 
the Fe--AnU considered. Then,  the  surface  under  the  Fe--AnU  is 
computed in between the two limits of the observed spectrum. This is 
a kind of ``observed EW''.  The  same  is  done  for  the  synthetic
spectrum. If the two surfaces are not equal,  the  Fe  abundance  is 
modified and a new synthetic spectrum is computed, and so  on  until 
the computed  surface  matches  the  observed one.  The  adopted  Fe 
abundance is the average of the values found from all the    Fe--AnU 
considered.

\subsubsection{Determination    of  the  atmospheric parameters
$\epsilon_{\rm turb}$, $T_{\rm eff}$, and log $g$}

\begin{figure}
\centering  \rotatebox{270}{
\includegraphics[width=6.0cm]{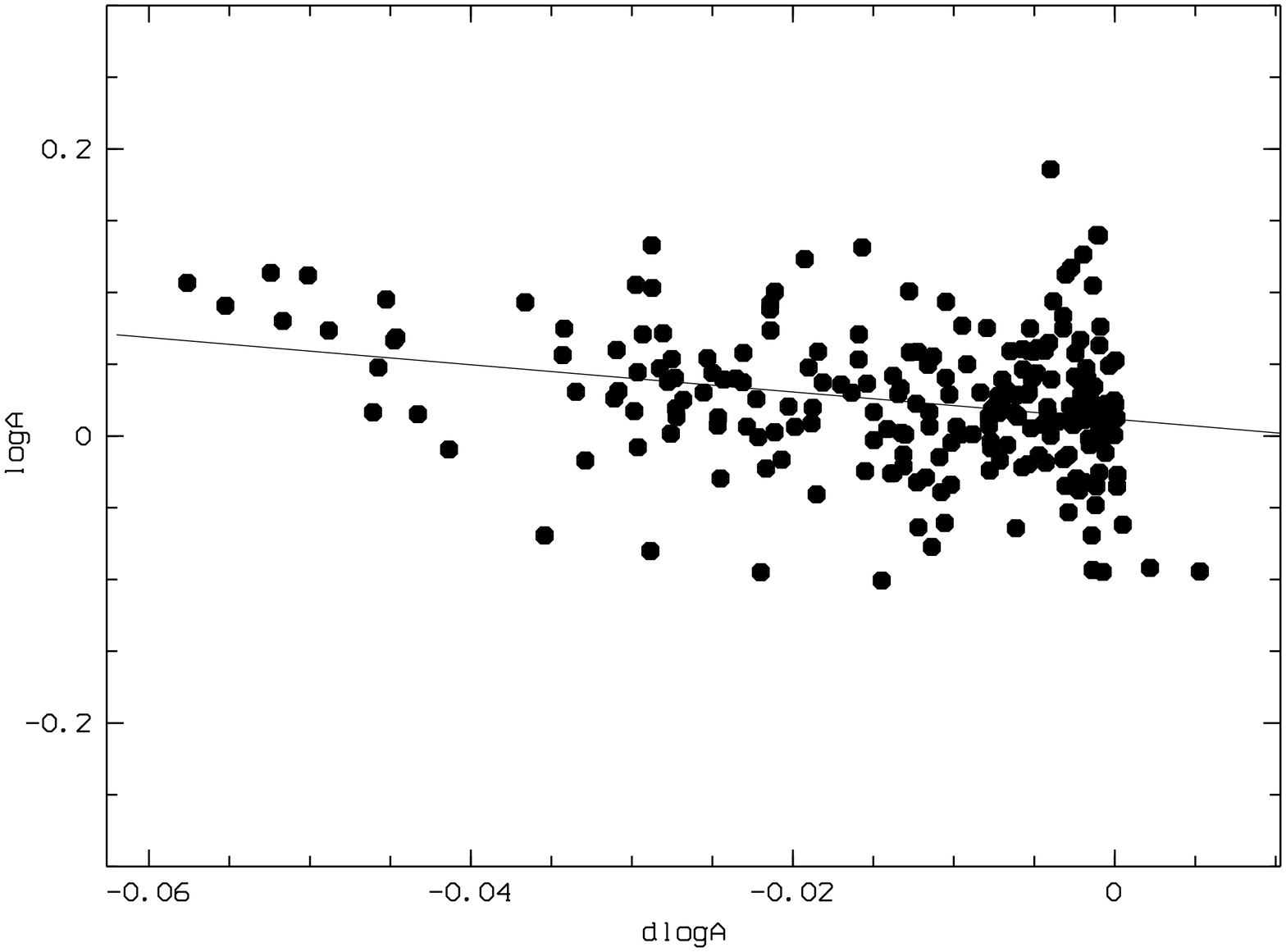}\includegraphics[width=6.0cm]
{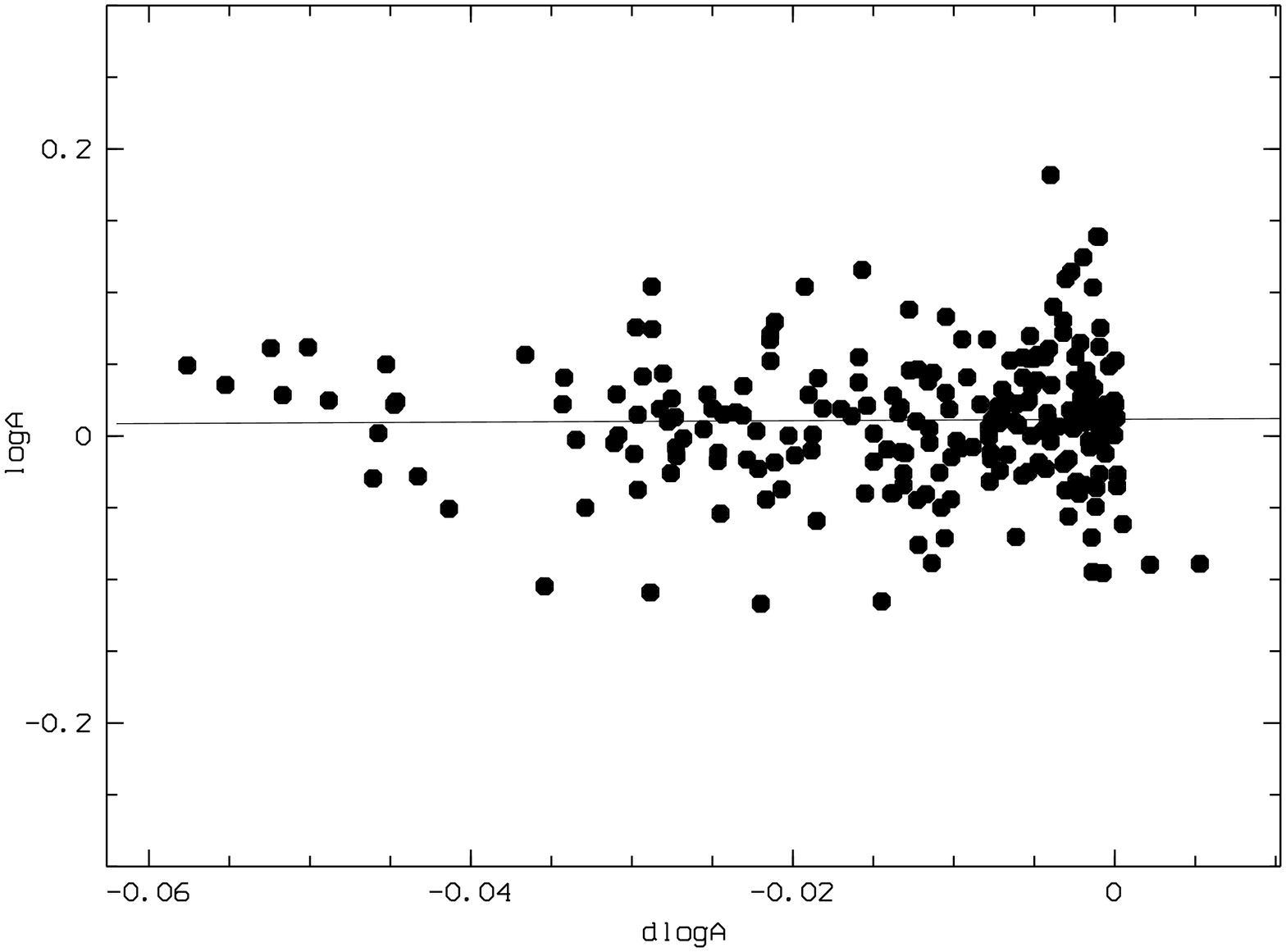}}
\caption{ Plot of the abundance deduced from each AnU  as a 
function of the sensitivity of the AnU to $\epsilon_{\rm turb}$ 
, for   $\epsilon_{\rm turb}$   =  1.20  km s$^{-1}$ 
(\emph{top}) and for $\epsilon_{\rm turb}$   =  1.30  km s$^{-1}$ 
(\emph{bottom}).  
The  slopes obtained by linear regression  are -0.96 $\pm$ 0.24 
for $\epsilon_{\rm turb}$   =  1.20  km s$^{-1}$ (initial guess) and 
-0.04 $\pm$ 0.24 for $\epsilon_{\rm turb}$   =  1.30  km s$^{-1}$
 (adopted value).}
\label{fig:klm}
\end{figure}

 In a first step, the sensitivity of each Fe--AnU to the atmospheric  
parameter
considered  is  determined.  First, we deal  with the microturbulence 
velocity   $\epsilon_{\rm turb}$.   We    adopt   a    value      of 
$\epsilon_{\rm turb}$ and compute the Fe abundance from each Fe--AnU. 
Then, we change the value of $\epsilon_{\rm turb}$ by a fixed amount 
and recompute the Fe abundances using this new model. The difference 
in Fe abundance obtained when  changing  $\epsilon_{\rm turb}$   
gives the sensitivity of each Fe--AnU to  this  parameter.  Once  the 
sensitivity of each Fe--AnU to $\epsilon_{\rm turb}$ is  determined, 
a diagram of    abundance  versus  sensitivity  is  constructed  and
$\epsilon_{\rm turb}$   is   determined   in  order  to  remove  any 
correlation: the  deduced   abundance   should  not  depend  on  the 
sensitivity to $\epsilon_{\rm turb}$. This procedure  is  similar to 
the classical method where $\epsilon_{\rm turb}$ is adjusted so that 
the abundance does not correlate with the line EW (the latter roughly
giving the sensitivity to $\epsilon_{\rm turb}$  for weak to medium 
strength lines). The microturbulence   is  determined  in  order  to 
remove any correlation between the abundances and  the sensitivities  
(see Figure~\ref{fig:klm}). In simple words, we want all the   Fe--AnU 
to  give us the    same      global    abundance,   whatever   their 
sensitivity to $\epsilon_{\rm turb}$. 

Following the same method, the correction to  the  abundances of our
Fe--AnU is computed for two models   differing  by  their   value of
$T_{\rm eff}$. Once the sensitivity of our Fe--AnU to  $T_{\rm eff}$ 
is computed, $T_{\rm eff}$ is   adjusted  to remove any  correlation    
between  the abundances and the sensitivities to $T_{\rm eff}$, which 
is similar to removing  the  trend  in  the  abundance  versus  line 
excitation potential plot. 

The same method is applied  again to determine log $g$; i.e. the 
sensitivity to log $g$ of the Fe--AnU  is  computed, and the  surface  
gravity is adjusted to remove any correlation between  the  abundances  
and  the sensitivities to log $g$.

A   complete  cycle  includes  determinations  of  $\Omega$,  [Fe/H],
 $\epsilon_{\rm turb}$, $T_{\rm eff}$, and log $g$. The iteration 
ends after convergence, i.e. when the corrections on  all  parameters
become much smaller  than  the error bars.

\subsubsection{Determination of $v$ sin$i$}

Once convergence  is  reached,  $v$ sin$i$   is determined by 
adjusting the synthetic spectrum by an $\chi^2$ minimization for
every Fe--AnU, and the average value is  taken  as  the  final  
$v$ sin$i$. If this final value    significantly  differs  from  the 
initial value, the whole iteration process is repeated.  As mentioned
above, $v$ sin$i$ is not adjusted during the iteration cycles  because 
it is strongly anticorrelated  with $\Omega$. Adjusting both parameters 
together thus produces instabilities.

\subsection{Computation of the error bars on the parameters}

For each parameter,   we must distinguish the intrinsic uncertainty 
and the  uncertainties coming from errors on the other    parameters 
(external uncertainties). 

\subsubsection{Intrinsic uncertainties}

For $T_{\rm eff}$, log $g$, and $\epsilon_{\rm turb}$,  the  standard 
deviation on  the  slope  in the  sensitivity   versus 
abundance  of Fe--AnU  diagram allows us to compute the intrinsic   uncertainty  
on these  parameters, i.e.   the    statistical   deviation.  As  an 
example,   the  uncertainty   on  the  slopes  in  Figure~\ref{fig:klm} 
is 0.24, and the difference between the two slopes  
is 1.00, for a   difference  in   $\epsilon_{\rm turb}$  =  0.10 km 
s$^{-1}$  between   the   models. The   intrinsic   uncertainty   on 
$\epsilon_{\rm turb}$ will be  in this case: 
\begin{equation}
\sigma = 0.10/1.00\times 0.24 = 0.02 \textrm{ km s$^{-1}$}.
\end{equation}

The convolution parameters $v$ sin$i$ and $\Omega$,  as well as
the iron abundance [Fe/H],  are 
determined independently for each Fe--AnU.  The    adopted    value   
is the average value for all the  Fe--AnU  used,  and  the  intrinsic 
uncertainty is the standard deviation  of the mean.

\subsubsection{Uncertainty on one parameter  due   to  the  intrinsic 
uncertainty on another parameter}

To compute the uncertainties on, e.g., log $g$ coming from another 
parameter, e.g., $T_{\rm eff}$,  we make a new determination of 
log $g$ using  a 
model with $T_{\rm eff}$ differing from the  adopted   value   by its 
intrinsic uncertainty. As an example, if we have a final value for
 log  $g$ =  4.34
dex and a final value for $T_{\rm eff}$ = 6400 K with  a   standard
deviation   =  50  K,  we set  $T_{\rm eff}$ to 6450 K  (or  6350 K)  
and determine log $g$. The difference between the obtained value  for 
log $g$ and 4.34 dex is the uncertainty on  log $g$  coming  from the 
uncertainty on $T_{\rm eff}$.

For each atmospheric parameter, we  compute the uncertainties coming 
from all the other parameters. The total error bar is then   computed 
as a quadratic sum of these uncertainties.  The intrinsic uncertainty 
is added  quadratically to the external  uncertainties  to obtain the 
final error bar.  An  exception  exists  for $v$  sin$i$,  for  which  we 
consider the intrinsic uncertainty and the  uncertainty  coming  from  
$\Omega$ as fully correlated. In this case, the intrinsic uncertainty   
is added linearly to the  uncertainty  coming  from $\Omega$, and the 
result is added quadratically to the  other  uncertainties  to obtain 
the final error bar on $v$  sin$i$.

\subsection{Test of the APASS method on an ELODIE spectrum of the Moon}

As a test, we determined the  parameters  of  the Sun  using   an
ELODIE spectrum of the sunlight reflected by the Moon.    Starting 
with $T_{\rm eff} =   6000$ K,  log $g = 4.00$ , $\epsilon_{\rm turb} 
=  2.00$, [Fe/H] = 1.00, and $v$  sin$i = 5$    km   s$^{-1}$   as 
initial guess parameters,  the method converged to the following  
parameters: $T_{\rm eff} =   5786 \pm 25 $ K,  log $g = 4.45  \pm 
0.03 $,  $\epsilon_{\rm turb} =  1.01  \pm 0.06$ km s$^{-1}$ , [Fe/H] 
= $+0.01 \pm 0.02 $, and $v$  sin$i = 1.94 \pm 0.72$  km  s$^{-1}$.
These are  very close  (and   well  within  error  bars)  to  the  ``expected'' 
solution, i.e. the   set   of   values   used  in  our  solar  model: 
$T_{\rm eff} =   5778$ K, log $g = 4.44$, $\epsilon_{\rm turb}= 1.00$  
km s$^{-1}$, [Fe/H] = 0.00 , $v$ sin$i = 2.00 $ km s$^{-1}$.    
This test demonstrates the   correct  convergence  of   APASS and the 
consistency of the ELODIE spectrum with the Jungraujoch Solar Atlas.

\subsection{Abundance determination}

Once we have obtained the atmospheric parameters, we determine   the 
atmospheric abundances  for  every  element  for  which we have  AnU 
containing significant lines of this element. We proceed by   steps, 
in order to avoid accumulating the  uncertainties. We  first  adjust 
the abundance of elements within  AnU  containing  lines  of  these 
elements alone or blended only with iron lines.  For  the  sake  of 
clarity, let us call A such an element. Once we have determined  the 
Fe abundance and the abundance of A, we can  proceed  determining
the abundance of an element X whose lines are blended with lines  of 
Fe and A. And so on, until all elements with  significant  lines  are 
included. 

\subsection{Computation of the error bars on the abundances}

The  uncertainty  is computed taking into account the part    coming 
from the  uncertainty on  the abundances of    the  other   elements 
included  in the AnU that we used, along with the uncertainty   on  the   
parameters. Let us consider an example. We   determine  [Si/H] 
using ten lines, contained in 8       different  AnU.  The  standard 
deviation of the mean is 0.02 dex. Eight of these lines are  blended 
in their AnU with Fe and Cr lines. We quadratically sum 0.02  dex to 
the intrinsic dispersion on  [Cr/H] and [Fe/H],  obtaining e.g.  0.04  
dex.  Then with different atmospheric  models we determine    the  
uncertainties  on [Si/H] coming  from the  intrinsic   uncertainties  
on  the  atmospheric parameters; i.e. we redetermine [Si/H]  with  a  
model  whose   $T_{\rm eff}$ equal  the   adopted  value  plus  or 
minus   the  intrinsic  uncertainty on $T_{\rm eff}$, and we do  the  
same  for    log $g$,  $\epsilon_{\rm turb}$, $\Omega$,  and  $v$ sin$i$.  
We   then  quadratically sum  all   the  uncertainties  on 
[Si/H], obtaining the final error bar. For elements for which we  do 
not have enough AnU to get a significant standard deviation  of 
the mean, we  take  the average of standard deviations for the  other 
elements except iron. We set to 4 the minimum number of AnU   needed   to 
judge the standard deviation significant.

%
%
\section{Results - comparison}

\subsection{Stellar parameters}

The final stellar parameters are presented in Table~4.  The  results 
are compared to  other determinations: \begin{itemize}
\item We  used the software TEMPLLOG (\cite{Rogers}) and the 
Str\"omgren indices extracted  from the Gaudi database (Table~3) to 
determine the atmospheric parameters $T_{\rm eff}$,  log $g$,    and  
[Fe/H],    with  a precision  around 200-250 K, 0.3 dex, and 0.2 dex, 
respectively, according to Rogers.  For HD 52265, 
the H$_{\beta}$ index  was  not  available on Gaudi, so we took  it  
from Hauck \& Mermilliod (1998).
\item For nine of the stars, $T_{\rm eff}$ was determined with 
the IR photometry method developed by Ribas et al.  (\cite{Ribas}), 
considered as more reliable than TEMPLOGG   by  the  COROT  team (C.  
Catala, private communication). 
\item For the stars HD 43587, HD 43318, HD 
45067, HD 57006, and HD  49933, which  have a low 
enough rotation velocity, we also used  our  equivalent--width 
method to analyse them. This method is described in  Bruntt et   al. 
(2004).
\item The star HD 52265 has a massive planet  orbiting  it      
that was detected by the radial velocity  method  (\cite{Butler}).   
Santos et al. derived stellar parameters  for   planet-host   stars  
using an    equivalent--width   method   (\cite{Santos}).   For  
HD 52265,  they made  two analyses, one based on a CORALIE 
spectrum and the other  on a FEROS spectrum.
\item Nine of the stars analysed in this work were also analysed by  
Bruntt using the VWA   software  (\cite{Bruntt},   \cite{Bruntt2}). 
This method is based, as is APASS, on the comparison  of  the  stellar     
spectrum with synthetic spectra.
\end{itemize} Figure 3 shows the   comparison  of   our  values for 
$T_{\rm eff}$,  log $g$, and [Fe/H]   with  the values obtained with 
TEMPLOGG and  our equivalent--width  method, and with the values 
obtained by Bruntt with VWA, while Figure 4 shows the comparison of
our values for $T_{\rm eff}$ with     the  values  obtained with the IR
photometry method.

\begin{table}
\label{tab:hgy}
\centering
\caption{Str\"omgren photometric indices of the COROT  main targets 
analysed in this work taken from the GAUDI database. The H$_{\beta}$ index 
for HD 52265 is taken from Hauck \& Mermilliod (1998)}
\begin{tabular}{cccccc}
\hline\hline  Star & V &    H$_{\beta}$ &   (b-y)   &  m1  &    c1\\  
\hline
HD 52265  &  6.294 & 2.619 & 0.360 & 0.186 & 0.423\\  
HD 43587  &  5.720 & 2.633 & 0.382 & 0.188 & 0.366\\
HD 43318  &  5.629 & 2.656 & 0.320 & 0.154 & 0.462\\
HD 45067  &  5.880 & 2.638 & 0.364 & 0.165 & 0.405\\
HD 57006  &  5.916 & 2.634 & 0.336 & 0.181 & 0.494\\
HD 49933  &  5.782 & 2.659 & 0.268 & 0.133 & 0.466\\
HD 175726 &  6.731 & 2.621 & 0.366 & 0.176 & 0.312\\
HD 177552 &  6.510 & 2.675 & 0.243 & 0.132 & 0.545\\
HD 184663 &  6.430 & 2.675 & 0.278 & 0.149 & 0.466\\
HD 46558  &  6.887 & 2.736 & 0.244 & 0.143 & 0.542\\
HD 171834 &  5.416 & 2.675 & 0.250 & 0.148 & 0.543\\
HD 49434  &  5.753 & 2.759 & 0.170 & 0.181 & 0.705\\
HD 55057  &  5.463 & 2.742 & 0.184 & 0.182 & 0.877\\
\hline
\end{tabular}
\end{table}

\begin{table*}
\label{tab:kop}
\centering
\caption{COROT targets and stellar parameters  obtained with APASS.
HD 49933 was analysed with an ELODIE spectrum (E) and    a  
high  $S/N$ HARPS spectrum (H). See text for   more   details   and  
Table 5  for references.}
\begin{tabular}{ccccccc}
\hline\hline Star &  $v$ sin$i$ (km/s)  &    $\epsilon_{\rm turb}$ 
(km s$^{-1}$)& $T_{\rm eff}$ (K) & log $g$ (dex)  &  [Fe/H]  (dex) 
& Ref. \\
\hline
{\bf HD 52265} &{\bf $5.0 \pm 1.3$} &{\bf $1.30 \pm 0.04$}&
{\bf $6179 \pm 18$}&{\bf $4.36 \pm 0.03$}& {\bf $+0.24 \pm 0.02$} & 
{\bf 1}\\ 
& &$1.33 \pm 0.08$&$6131 \pm 47$&$4.35 \pm 0.13$&$+0.25 \pm 0.06$& 3\\
& &$1.38 \pm 0.09$ &$6076 \pm 57$&$4.20 \pm 0.17$&$+0.20 \pm 0.07$& 4\\
& & &$6135 \pm 200$&$4.09 \pm 0.20$&$+0.05 \pm 0.20$ & 5\\
& & &$6136 \pm 51$& & & 7\\
{\bf HD 43587} &{\bf $5.8 \pm 0.9$}&{\bf $1.18 \pm 0.04$} & 
{\bf $5978 \pm 18$}&{\bf$4.29 \pm 0.02$} & {\bf$+0.02 \pm 0.02$}& 
{\bf 1}\\ 
& & &$5870 \pm 60$&$4.20 \pm 0.15$&$-0.09 \pm 0.11$&6\\
& &$2.13 \pm 0.10$&$5867 \pm 20$&$4.29 \pm 0.06$&$-0.13 \pm 0.02$&2a\\
& & &$6327 \pm 200$&$4.54 \pm 0.20$&$+0.24  \pm 0.20$&5\\
& & &$5925 \pm 65$& & &7\\
{\bf HD 43318}&{\bf $7.6 \pm 2.2$}&{\bf $1.56 \pm 0.08$}&
{\bf $6196 \pm 33$}&{\bf $3.68 \pm 0.06$}&{\bf $-0.16 \pm 0.03$} & 
{\bf 1}\\ 
& & &$6190 \pm 90$&$3.70 \pm 0.15$&$-0.21 \pm 0.11$&6\\
& &$2.13 \pm 0.31$ &$6045 \pm 40$&$3.63 \pm 0.07$&$-0.24 \pm 0.03$&2a\\
& & &$6517 \pm 200$&$4.25 \pm 0.20$&$-0.06  \pm 0.20$&5\\
& & &$6220 \pm 98$& & &7\\
{\bf  HD 45067} &{\bf $7.9 \pm 1.3$}&{\bf $1.31 \pm 0.05$} &
{\bf $6070 \pm 23$}&{\bf $3.99 \pm 0.04$} &{\bf $-0.02 \pm 0.02$}& 
{\bf 1}\\ 
& & &$5970 \pm 100$&$3.80 \pm 0.15$&$-0.17 \pm 0.11$&6\\
& &$2.00 \pm 0.13$ &$5913 \pm 30$&$3.91 \pm 0.08$&$-0.15 \pm 0.02$&2a\\
& & &$6358 \pm 200$&$4.37 \pm 0.20$&$+0.02  \pm 0.20$&5\\
& & &$6092 \pm 59$& & &7\\
{\bf  HD 57006} &{\bf $8.2 \pm 1.5$}&{\bf $1.68 \pm 0.06$} & 
{\bf $6177 \pm 26$} & {\bf $3.66 \pm 0.05$} & {\bf $-0.02 \pm 0.02$}& 
{\bf 1}\\ 
& & &$6180 \pm 70$&$3.60 \pm 0.15$&$-0.08 \pm 0.11$&6\\
& &$2.72 \pm 0.09$&$6026 \pm 22$&$3.61 \pm 0.05$&$-0.09 \pm 0.02$&2a\\
& & &$6250 \pm 200$&$3.74 \pm 0.20$&$+0.06  \pm 0.20$&5\\
{\bf  HD 49933} & {\bf $10.3 \pm 1.8$}&{\bf $1.65 \pm 0.10$}&
{\bf $6700 \pm 65$} & {\bf $4.25 \pm 0.09$} & {\bf $-0.34 \pm 0.03$}&
{\bf 1 (E)}\\
& {\bf $10.0 \pm 1.6$}  & {\bf $1.84 \pm 0.08$}  & {\bf $6735 \pm 53$} & 
{\bf $4.26 \pm 0.08$} &  {\bf $-0.37 \pm 0.03$}&{\bf 1 (H)}\\ 
& & &$6780 \pm 70$&$4.30 \pm 0.20$&$-0.30 \pm 0.11$&6\\
& &$2.53 \pm 0.31$ &$6467 \pm 56$&$4.27 \pm 0.08$ & $-0.37 \pm 0.04$&2a
(E)\\
& &$2.12 \pm 0.24$ &$6539 \pm 47$&$4.41 \pm 0.07$ & $-0.45 \pm 0.02$&2a 
(H)\\
& &$2.33 \pm 0.24$ &$6538 \pm 44$&$4.38 \pm 0.08$ & $-0.47 \pm 0.02$&2b 
(H)\\
& & &$6543 \pm 200$&$4.24 \pm 0.20$&$-0.38  \pm 0.20$&5\\
& & &$6747 \pm 64$& & &7\\
{\bf HD 175726} & {\bf $12.7 \pm 1.7$} & {\bf $1.32 \pm 0.08$} & 
{\bf $6217 \pm 32$} & {\bf $4.61 \pm 0.04$} & {\bf $+0.03 \pm 0.03$}& 
{\bf 1}\\ 
& & &$6227 \pm 200$&$4.73 \pm 0.20$&$-0.05  \pm 0.20$&5\\
{\bf  HD 177552} & {\bf $38.9 \pm 2.5$} & {\bf $1.88 \pm 0.10$} & 
{\bf $7025 \pm 55$} & {\bf $4.45 \pm 0.08$} &{\bf $-0.03 \pm 0.03$} &
{\bf 1}\\ 
& & &$6657 \pm 200$&$4.03 \pm 0.20$&$-0.38  \pm 0.20$&5\\
& & &$7042 \pm 69$& & &7\\
{\bf HD 184663} & {\bf $57.3 \pm 5.3$} & {\bf $2.01 \pm 0.26$} & 
{\bf $6511 \pm 108$} & {\bf $4.28 \pm 0.21$} & {\bf $-0.13 \pm 0.06$} &
{\bf 1}\\
& & &$6600 \pm 90$&$4.50 \pm 0.50$&$-0.19 \pm 0.21$&6\\
& & &$6700 \pm 200$&$4.40 \pm 0.20$&$-0.13  \pm 0.20$&5\\
{\bf HD 46558} & {\bf $60.0 \pm 5.9$} & {\bf $1.63 \pm 0.21$} & 
{\bf $7053 \pm 92$} & {\bf $4.44 \pm 0.14$} & {\bf $+0.04 \pm 0.07$} &
{\bf 1}\\ 
 & & &$7223 \pm 200$&$4.71 \pm 0.20$&$-0.24  \pm 0.20$&5\\
 & & &$6922 \pm 66$& & &7\\
{\bf HD 171834} & { \bf $70.0 \pm 6.3$} & {\bf $1.64 \pm 0.21$}& 
{\bf $6833 \pm 105$} & {\bf $4.14 \pm 0.15$}  &{\bf $+0.01 \pm 0.07$} & 
{\bf 1}\\
 & & &$6840 \pm 200$&$4.60 \pm 0.50$&$-0.25 \pm 0.21$&6\\
 & & &$6658 \pm 200$&$4.04 \pm 0.20$&$-0.17  \pm 0.20$&5\\
 & & &$6879 \pm 67$& & &7\\
{\bf HD 49434} & { \bf $85.4 \pm 6.6$} & {\bf $2.77 \pm 0.26$} & 
{\bf $7632 \pm 126$} & {\bf $4.43 \pm 0.20$} & {\bf $+0.09 \pm 0.07$} & 
{\bf 1}\\ 
& & &$7300 \pm 200$&$4.40 \pm 0.45$&$-0.04 \pm 0.21$&6\\
& & &$7343 \pm 200$&$4.21 \pm 0.20$&$+0.01  \pm 0.20$&5\\
& & &$7346 \pm 69$& & &7\\
{\bf HD 55057} & {\bf $130.1 \pm 7.5$} & {\bf $3.20 \pm 0.27$} & 
{\bf $7478 \pm 105$} & {\bf $3.84 \pm 0.17$} & {\bf $+0.26 \pm 0.08$} & 
{\bf 1}\\
& & &$7580 \pm 250$&$3.60 \pm 0.50$&$+0.14 \pm 0.21$&6\\
& & &$7141 \pm 200$&$3.47 \pm 0.20$&$+0.08  \pm 0.20$&5\\
\hline
\end{tabular}
\end{table*}

\begin{table*}
\centering
\caption{The references for Table 4}
\begin{tabular}{cc}
\hline
Ref.{\bf 1} & This study with APASS  \\
Ref. 2a & EW method,  Kurucz 1993 model  \\
Ref. 2b & EW method, CGM model  \\
Ref. 3 & Santos et al. (2004), EROS spectrum \\
Ref. 4 & Santos et al. (2004), CORALIE spectrum \\
Ref. 5 & TEMPLOGG using Str\" omgren indices from Gaudi database \\
Ref. 6 & Bruntt et al. (2004) using VWA \\
Ref. 7 & IR photometry (Ribas et al., 2003)\\
\hline
\end{tabular}
\end{table*}

\begin{figure*}
\label{fig:lpl}
\centering                     
\includegraphics[width=18.0cm]{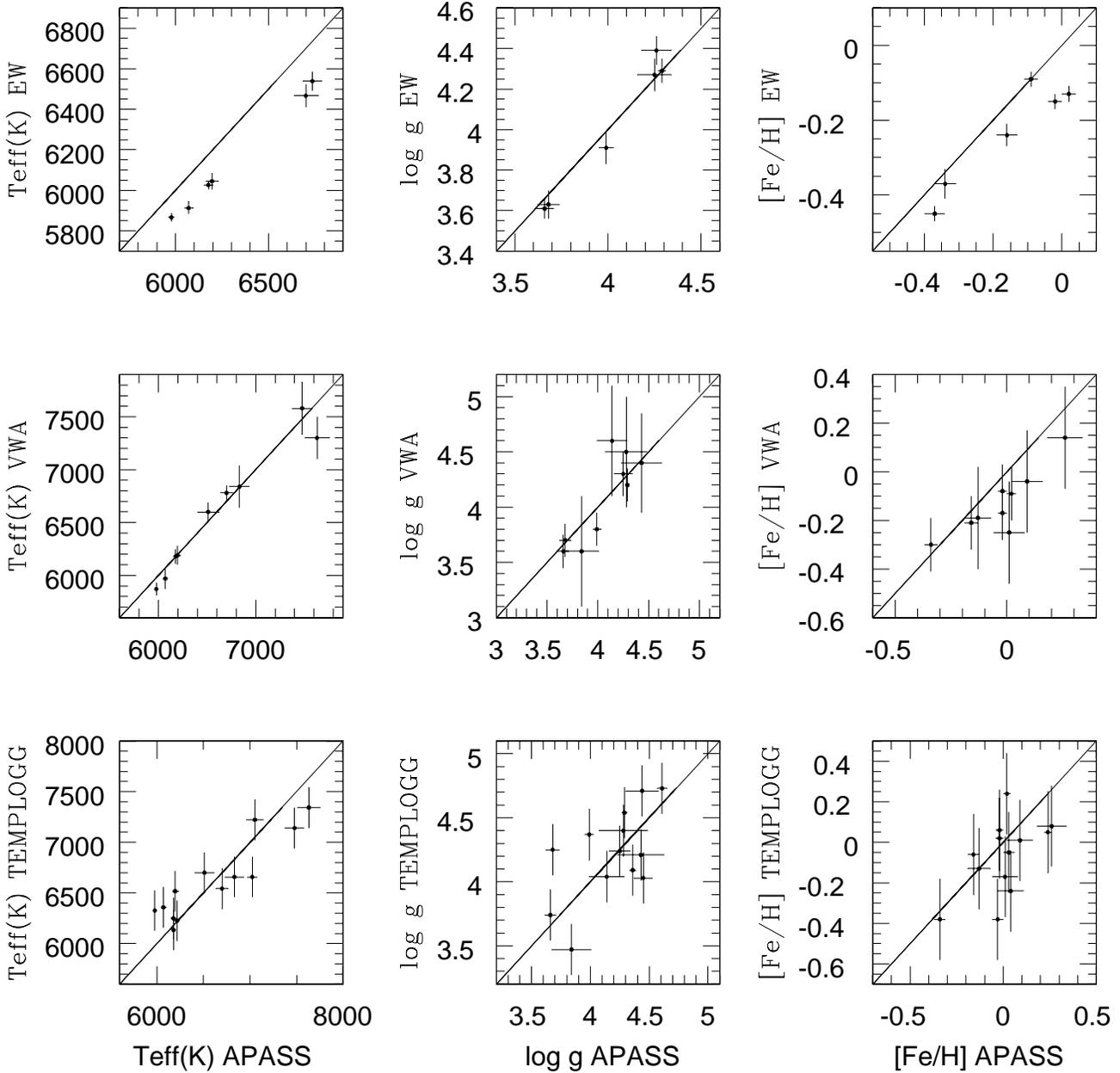}
\caption{Comparison of the atmospheric parameters $T_{\rm eff}$, log $g$ 
, and [Fe/H], derived in this    work with those obtained  with  our  EW  
method (\emph{top}), by  Bruntt  using  VWA (\emph{middle}),  and   with 
TEMPLOGG using Str\"omgren indices from the Gaudi database (\emph{bottom}).}
\end{figure*}

\subsubsection{Comparison with TEMPLOGG values}

For $T_{\rm eff}$, we found an average difference  $<\delta >$ 
= -2 K when comparing our values with those obtained with TEMPLOGG.
Nevertheless, as can be seen   in  Figure 3,  the  dispersion  is  high 
($\sigma = 253$ K). Despite the small  number  of  points,  a  tendency 
seems to appear in the plot. TEMPLOGG tends to  give higher values than 
APASS  for $T_{\rm eff}\sim$  6000  K,  and  lower  values  for  higher 
$T_{\rm eff}$. For log $g$, $<\delta > =  -0.03$ dex   with  $\sigma$  
=  0.30  dex, and for [Fe/H] the $<\delta > = +0.07$ dex and $\sigma =  
0.18$ dex. We thus conclude that our results generally   agree
with the values obtained with TEMPLOGG. However, the scatter is rather
large, and can generally be explained by the larger error bars in the 
photometric TEMPLOGG determination.  

\subsubsection{Comparison with IR photometry values}

As can be seen from Figure 4, agreement  between  the   $T_{\rm eff}$ 
obtained with APASS and from IR photometry is  very  good,  within  the 
small error bars for  most  of  the  stars: $<\delta >  = + 40$ K, with 
$\sigma = 110$  K.  Only  the hottest  star, HD 49434, shows  
significant disagreement. If we do  not  consider it,   $<\delta >$  is 
reduced to +9 K and $\sigma$ to 55 K.

\begin{figure}
\label{fig:lmm}    
\centering                   
\includegraphics[width=8.0cm]{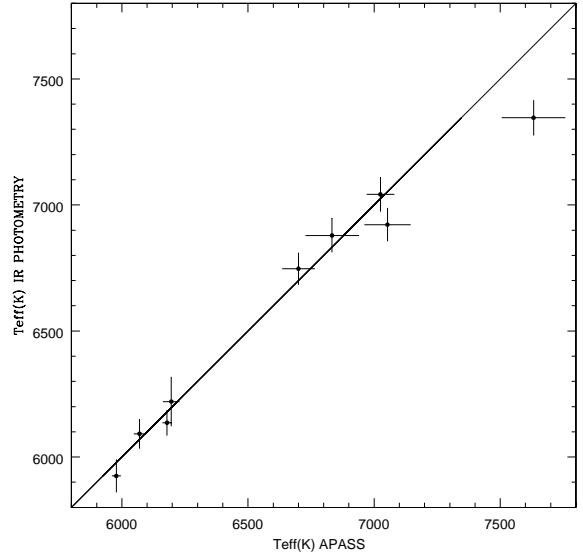}
\caption{Comparison of the $T_{\rm eff}$ derived in this work with those 
obtained with the IR photometry method of Ribas et al. (2003).}
\end{figure}

\subsubsection{Comparison with the values obtained with our EW method}

We find $<\delta > = +167$ K with  $\sigma = 187$ K for  $T_{\rm eff}$,
when comparing our results using APASS with  those obtained    using
our EW method. The values obtained   for [Fe/H] ($<\delta > = + 0.09$  
dex, $\sigma = 0.11$ dex)  and   for  $\epsilon_{\rm turb}$ ($<\delta > 
= -0.75$ km/s, $\sigma  =  0.86$ km/s) are not in very  good  agreement 
either. Only the values obtained for log $g$ are consistent ($<\delta > 
= 0.00$ dex, $\sigma  = 0.08$ dex). 

 Our  EW method uses Gaussian and Voigt profile--fitting to measure  the 
EW of single lines. All the stars  analysed  here  have  a  $v$  sin$i$ 
higher than the Sun, and we wanted to  know  if  the  measured  EW  are 
 independent of the rotational broadening.  We 
measured the EW of three  synthetic  FeI  lines computed with different 
convolution parameters $v$ sin$i$, with solar values for $\epsilon_{\rm 
turb}$, $T_{\rm eff}$, log $g$, and [Fe/H] and with  a  fixed  value  of 
$\Omega$. We  also computed   the theorical  equivalent--width  of  
these  lines.  Our  results in Figure 5 show that
the measured EW is sensitive to the  rotational broadening of the lines. 
The fit of Gaussian profiles tends to 
give EW values that are too low for  low $v$ sin$i$, and  the measured 
EW  tends  to  increase with  $v$ sin$i$. The fit of Voigt profiles  
gave   good results for low values of $v$ sin$i$,   but seemed  to 
give absolutely incorrect  values  for  values of  $v$  sin$i$  higher 
than 4 km/s.
The same test with a fixed value of  $v$ sin$i$ and different values of
$\Omega$ showed also a clear dependence  of  the   measured EW  on  the 
convolution magnitude of the spectrum.

We plot in  Figure 6  the observed  differences between the  two  methods 
as a function of $v$ sin$i$. For $T_{\rm eff}$,  a  clear  tendency  
appears. The higher the rotational broadening of the   spectrum, 
the larger the   discrepancy between the 2 methods,  with the   EW
method giving values lower than  APASS. For the other parameters,   we  
can  only  say  that  lower  [Fe/H] and  higher   $\epsilon_{\rm turb}$ 
are found with  the  EW method.   All  those results  seem  to indicate  
that  an    EW  method   using Gaussian and Voigt profiles fitting
is not suitable for the analysis of stars  with a projected rotation
velocity that is significantly different from that of the Sun.

\begin{figure*}
\label{fig:ioi}
\centering                     
\includegraphics[width=14.0cm]{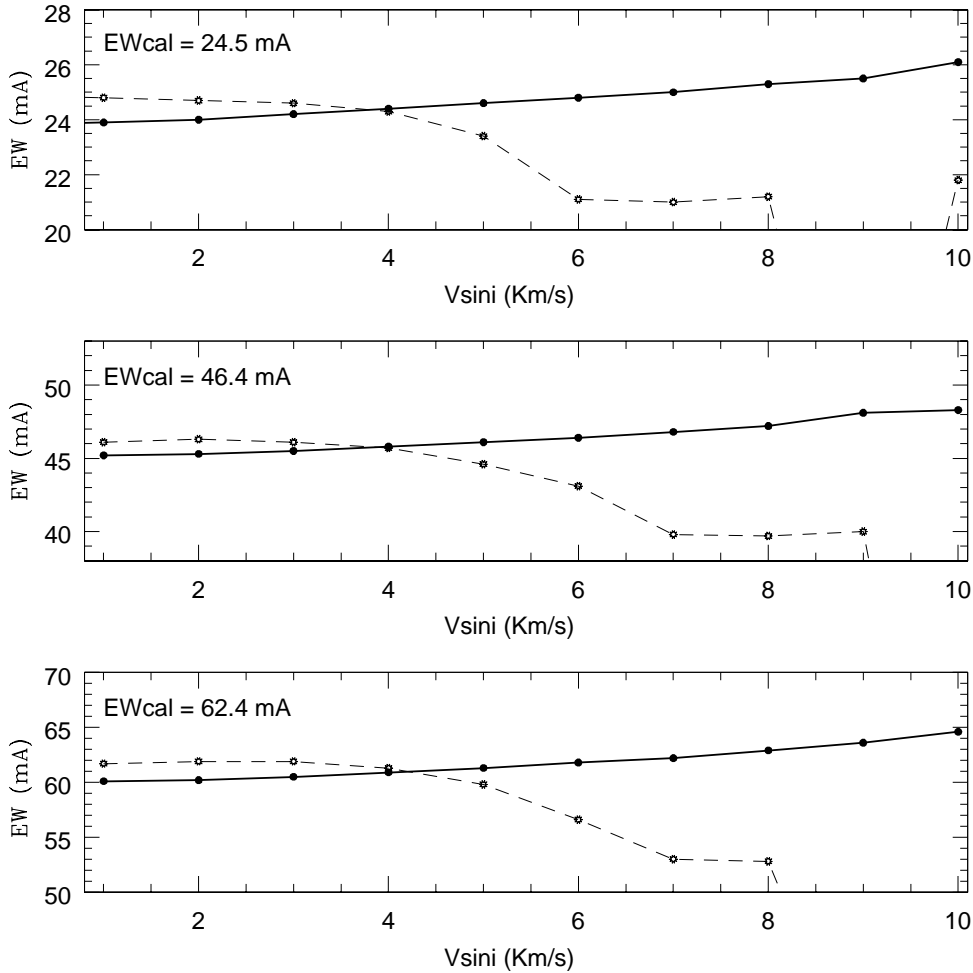}
\caption{Measured equivalent--widths  of  3  synthetic  FeI  lines  by 
Gaussian (solid line) and Voigt (dashed line)  profile--fitting,  for 
different values of $v$ sin$i$. $\Omega$ is  fixed to  2   km/s.   The 
theorical  value of  the  EW appears in the upper left corner of  each 
plot.} 
\end{figure*}

\begin{figure*}
\label{fig:yuo}
\centering                     
\includegraphics[width=12.0cm]{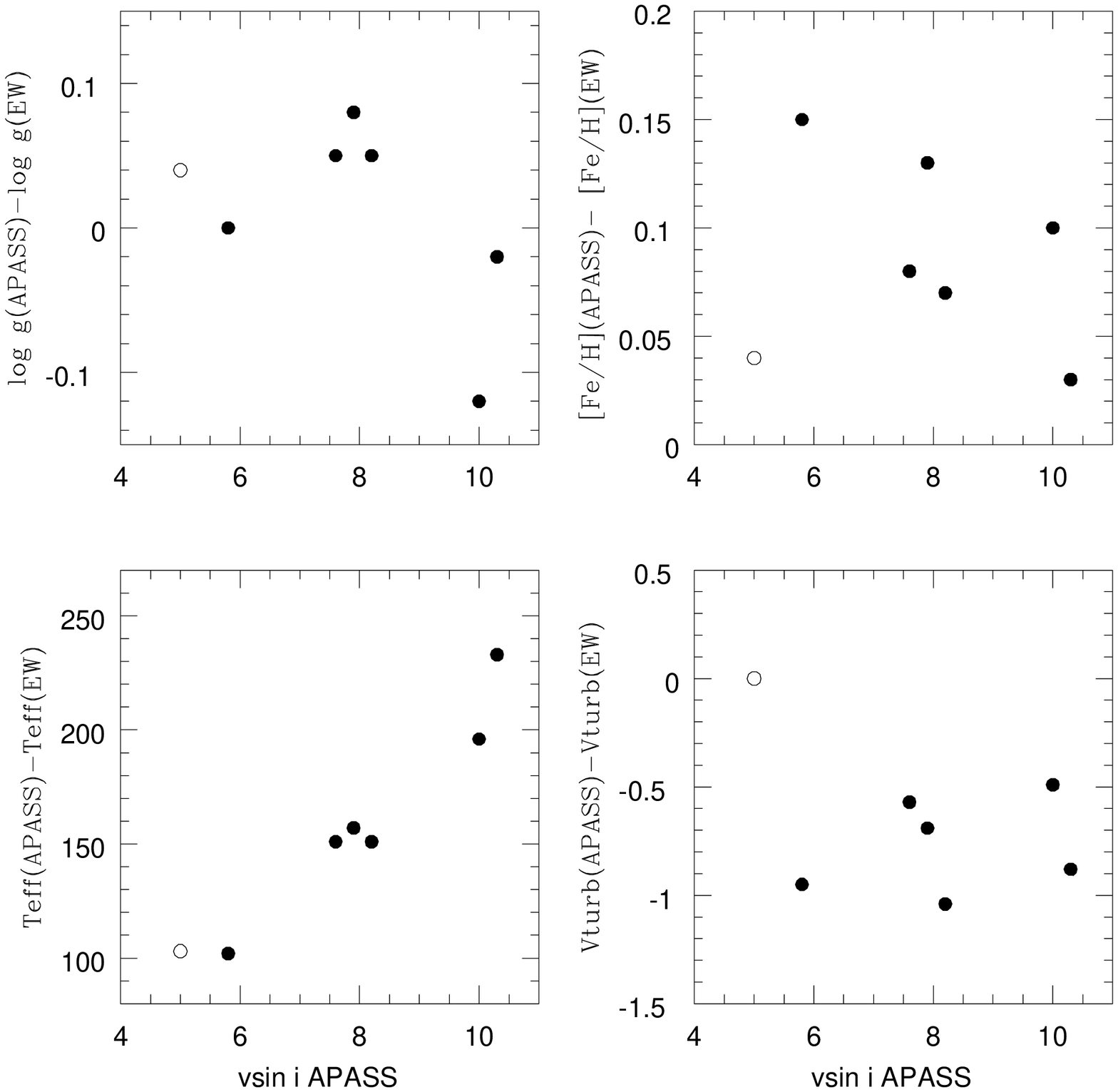}
\caption{Differences between  the values obtained with APASS and with
the EW method as a function  of $v$   sin$i$.  Open  symbol:  results 
obtained   by  Santos  et al.  for HD 52265 with  a  CORALIE 
spectrum.}
\end{figure*}

\subsubsection{Comparison with VWA values}

A small $<\delta >$ = + 29 K is found for $T_{\rm eff}$ ($\sigma   = 
140$ K), and $<\delta >$ = -0.02 dex for log $g$  with a  $\sigma $ =
 0.21 dex. 
For [Fe/H], there is  marginal disagreement, with $<\delta > = 
+0.10$ dex  and 
$\sigma  = 0.13$ dex. Despite  the  fact that these  two  methods use  
synthetic  spectra  to obtain  atmospheric parameters, our error bars 
are significantly smaller than those obtained with VWA. We believe 
that this could be due to four causes :
\begin{enumerate}
\item  strict selection  of  the  spectral   regions  used  for  the 
analysis,
\item  precise determination of the continuum,
\item  precise   determination  of  the  convolution  level  of  the 
spectrum,
\item the use of packs of    blended  lines  as  analysis  units, with 
all the log $gf$ of the main lines of the packs  precisely determined
from a high resolution solar atlas. This point is very important. If we 
would   restrict  our  analysis  to  isolated  lines, the  number  of 
independant data  available for  our analysis would  decrease drastically 
with the increase  of  the line broadening.   Using  blended 
lines allows us to extract  a   high   quantity  of  information  from 
spectra, even if they are moderately  broadened. 
\end{enumerate}

\subsubsection{HD 52265: comparison with results obtained  by 
Santos et al.(2004)}

 Table 4 shows that our results for  HD 52265 with APASS  
are  in  good agreement with those of Santos et al.  from 
the analysis   of their FEROS spectrum of this star, as can be seen  
from  Table~4. As our 
analysis is based on a CORALIE spectrum, it is more meaningful to compare 
our results with the CORALIE results of Santos et al. We find 
$\Delta T_{\rm eff}$ = 103 K. As can be seen from  Figure 6,  it 
agrees with the  overall tendency noticed when comparing  the   
 results  of  APASS   and of our EW method. A linear regression made on 
the  values 
of  $v$ sin$i$ as a function of  the  difference  in    $T_{\rm eff}$ 
gives an  intercept = 1.54 km s$^{-1}$, near the observed $v$ sin$i$  
of  the  Sun,  $1.9 \pm 0.3 $  km s$^{-1}$  (\cite{Soderblom}).   The 
correlation coefficient $r$ is  0.97. 
Our values for $\epsilon_{\rm turb}$, log $g$, and [Fe/H] are   within 
the error bars of Santos et al., but their values  are  not   inside   
our  own error bars. For these  parameters, no information comes from Figure 6,
except the fact that the EW method used by  Santos  et al.  does  not 
seem to give much higher $\epsilon_{\rm turb}$ than APASS, unlike our 
own EW method. We notice that Santos et al. measured  the  
equivalent--width by fitting a Gaussian profile, whatever the  strength  of  
 the  line. This method   will  tend  to  give  too  low an EW  for    strong  
lines, while $\epsilon_{\rm turb}$ is determined in  the  EW    method   by  
 cancelling the correlation between  the line abundances  and  their
 EW. An EW that is too low  for stronger lines will produce an underestimation 
of   $\epsilon_{\rm  turb}$, so the  fact  that  the  
$\epsilon_{\rm turb}$ found by Santos  et al. for   HD 52265 
is near   our  own  value  could   be a coincidence that comes from  the 
combined effects of ($i$) the systematic error due to  the  use of Gaussian 
profiles,  whatever the  strength  of the line, and of ($ii$) the systematic  
error  due  to  the  fact that the measurement of the EW by 
non-rotationally broadened profiles fitting is sensitive
to the degree to which the spectrum is convolved.

\begin{table*}
\centering
\caption{Abundances obtained with APASS for the  COROT  main targets.  
HD 49933 was analysed with an ELODIE spectrum (E) and a  high 
$S/N$ HARPS spectrum (H). See text for more details.}
\begin{tabular}{cccccccc}
\hline\hline&HD 52265&HD 43587 &HD 43318 &
HD 45067&HD 57006&HD 49933(E)&HD 49933(H)\\
\hline
\textrm{[Na/H]} &$+0.29 \pm 0.04$ & $+0.06 \pm 0.04$  &$-0.14 \pm 0.04$&
$+0.01 \pm 0.03$&$-0.02 \pm 0.04$&$-0.32 \pm 0.09$&$-0.35 \pm 0.05$\\ 
\textrm{[Si/H]}&$+0.30 \pm 0.03$ &  $+0.05 \pm 0.03$ & $-0.07 \pm 0.04$&
$+0.05 \pm 0.04$&$+0.12 \pm 0.05$&$-0.32 \pm 0.04$&$-0.33 \pm 0.04$\\ 
\textrm{[S/H]} &$+0.23 \pm 0.07$  & $-0.07 \pm 0.05$ & $-0.12 \pm 0.03$&
$+0.14 \pm 0.20$&$+0.02 \pm 0.18$&$-0.25 \pm 0.08$&$-0.31 \pm 0.05$\\ 
\textrm{[Ca/H]} &$+0.23 \pm 0.03$& $+0.02 \pm 0.03$  & $-0.12 \pm 0.04$&
$-0.04 \pm 0.03$&$-0.02 \pm 0.06$&$-0.43 \pm 0.04$&$-0.39 \pm 0.05$\\
\textrm{[Sc/H]} &$+0.26 \pm 0.04$& $+0.05 \pm 0.03$  & $-0.13 \pm 0.08$&
$-0.09 \pm 0.14$&$+0.14 \pm 0.15$&$-0.35 \pm 0.08$&$-0.34 \pm 0.10$\\ 
\textrm{[Ti/H]} &$+0.26 \pm 0.04$&$+0.08 \pm 0.05$   & $-0.11 \pm 0.08$&
$-0.12 \pm 0.08$&$-0.06 \pm 0.04$&$-0.29 \pm 0.04$&$-0.29 \pm 0.04$\\
\textrm{[V/H]} & $+0.22 \pm 0.04$ & $+0.07 \pm 0.05$ & $-0.21 \pm 0.09$&
$-0.10 \pm 0.07$&$+0.03 \pm 0.05$&$-0.43 \pm 0.11$&$-0.50 \pm 0.11$\\ 
\textrm{[Cr/H]} &$+0.27 \pm 0.03$&$-0.02 \pm 0.04$  &  $-0.21 \pm 0.05$&
$-0.09 \pm 0.04$&$-0.03 \pm 0.03$&$-0.37 \pm 0.05$&$-0.39 \pm 0.08$\\
\textrm{[Mn/H]} &$+0.27 \pm 0.03$ & $+0.01 \pm 0.04$ & $-0.25 \pm 0.05$&
$-0.11 \pm 0.08$&$-0.08 \pm 0.05$&$-0.29 \pm 0.10$&$-0.43 \pm 0.08$\\ 
\textrm{[Fe/H]} &$+0.24 \pm 0.02$ & $+0.02 \pm 0.02$ & $-0.16 \pm 0.03$&
$-0.02 \pm 0.02$&$-0.02 \pm 0.02$&$-0.34 \pm 0.03$&$-0.37 \pm 0.03$\\ 
\textrm{[Co/H]} &$+0.19 \pm 0.06$ & $-0.01 \pm 0.05$ & $-0.19 \pm 0.07$&
$-0.02 \pm 0.06$&$+0.02 \pm 0.07$&$-0.52 \pm 0.05$&$-0.53 \pm 0.07$\\ 
\textrm{[Ni/H]} &$+0.25 \pm 0.03$&$+0.02 \pm 0.03$  &  $-0.23 \pm 0.04$&
$-0.07 \pm 0.03$&$-0.04 \pm 0.04$&$-0.53 \pm 0.05$&$-0.51 \pm 0.05$\\
\textrm{[Cu/H]} &$+0.24 \pm 0.06$ & $-0.02 \pm 0.05$ & $-0.17 \pm 0.07$&
$-0.08 \pm 0.03$&$-0.10 \pm 0.08$&$-0.48 \pm 0.07$&$-0.46 \pm 0.08$\\ 
\textrm{[Y/H]} &$+0.31 \pm 0.08$ & $+0.03 \pm 0.03$  & $-0.16 \pm 0.05$&
$+0.00 \pm 0.07$&$+0.03 \pm 0.04$&$-0.21 \pm 0.08$&$-0.22 \pm 0.07$\\ 
\textrm{[Ba/H]} &$+0.26 \pm 0.06$&$-0.05 \pm 0.07$&$-0.03 \pm 0.08$&
$+0.10 \pm 0.11$&$+0.23 \pm 0.13$&$-0.17 \pm 0.09$&$-0.20 \pm 0.08$\\
\textrm{[Nd/H]} &$+0.22 \pm 0.06$ & $-0.21 \pm 0.08$ & $-0.08 \pm 0.06$&
$-0.10 \pm 0.12$&$+0.00 \pm 0.06$&$-0.25 \pm 0.10$&$-0.21 \pm 0.10$\\ 
\hline
\hline\hline&HD 175726&HD 177552&HD 184663&HD 46558&
HD 171834&HD 49434&HD 55057\\
\hline
\textrm{[Na/H]}&$-0.07 \pm 0.06$ & $-0.10 \pm 0.10$ &  $-0.14 \pm 0.09$&
$-0.06 \pm 0.15$&$-0.09 \pm 0.16$&$-0.07 \pm 0.18$&$+0.24 \pm 0.19$\\ 
\textrm{[Si/H]}&$+0.01 \pm 0.04$ & $+0.15 \pm 0.10$ &  $+0.09 \pm 0.14$&
$+0.15 \pm 0.10$&$+0.00 \pm 0.20$&$-0.12 \pm 0.15$&$+0.38 \pm 0.25$\\ 
\textrm{[S/H]}&$+0.10 \pm 0.08$  &  $+0.00 \pm 0.16$ & $+0.21 \pm 0.07$&
$-0.29 \pm 0.18$&$+0.01 \pm 0.15$&$+0.11 \pm 0.17$&\\ 
\textrm{[Ca/H]}&$+0.06 \pm 0.04$ &  $-0.19 \pm 0.09$ & $-0.17 \pm 0.09$&
$-0.05 \pm 0.09$&$-0.15 \pm 0.08$&$+0.03 \pm 0.10$&$+0.23 \pm 0.17$\\
\textrm{[Sc/H]}&$+0.10 \pm 0.07$&&&&&&\\ 
\textrm{[Ti/H]}&$+0.07 \pm 0.06$ & $+0.19 \pm 0.15$  & $-0.04 \pm 0.09$&
$+0.02 \pm 0.14$&$-0.06 \pm 0.10$&$+0.17 \pm 0.22$&$+0.43 \pm 0.31$\\
\textrm{[V/H]}&$+0.12 \pm 0.08$&&&&&&\\ 
\textrm{[Cr/H]}&$+0.03 \pm 0.04$  & $+0.00 \pm 0.04$ & $-0.11 \pm 0.07$&
$+0.01 \pm 0.09$&$-0.03 \pm 0.08$&$+0.07 \pm 0.09$&$+0.27 \pm 0.12$\\
\textrm{[Mn/H]}&$-0.01 \pm 0.04$  & $+0.10 \pm 0.12$ & $-0.16 \pm 0.15$&
$-0.07 \pm 0.12$&$+0.09 \pm 0.18$&&\\ 
\textrm{[Fe/H]} &$+0.03 \pm 0.03$ & $-0.03 \pm 0.03$ & $-0.13 \pm 0.06$&
$+0.04 \pm 0.07$&$+0.01 \pm 0.07$&$+0.09 \pm 0.07$&$+0.26 \pm 0.08$\\ 
\textrm{[Co/H]}&$-0.02 \pm 0.07$ & $-0.01 \pm 0.14$ &  $-0.10 \pm 0.13$&
$-0.08 \pm 0.18$&$+0.02 \pm 0.13$&$+0.13 \pm 0.15$&$+0.20 \pm 0.18$\\ 
\textrm{[Ni/H]}&$+0.01 \pm 0.04$  & $-0.08 \pm 0.10$ & $-0.19 \pm 0.14$&
$-0.02 \pm 0.17$&$-0.02 \pm 0.11$&$+0.06 \pm 0.08$&$+0.38 \pm 0.11$\\
\textrm{[Cu/H]}&$-0.10 \pm 0.08$&&&&&&\\ 
\textrm{[Y/H]}&$+0.16 \pm 0.05$&&&&&\\ 
\textrm{[Ba/H]}&$+0.28 \pm 0.08$&$+0.16 \pm 0.14$&$-0.26 \pm 0.13$&&&&\\
\textrm{[Nd/H]}&$+0.20 \pm 0.09$&&&&&&\\ 
\hline
\end{tabular}
\end{table*}

\subsection{Abundances}

We present the abundances obtained in this work in Table 7. These
are abundances relative to the Sun. From the abundance patterns we find 
no evidence of chemically peculiar stars.

%
\section{Conclusions}

We have developed a method able to precisely   determine the 
 atmospheric parameters and abundances  of stars with   
moderate to high  projected rotational velocity. This differential 
method,  APASS, is   based  on  the  plane-parallel  homogeneous  
atmosphere and LTE hypothesis.

With  APASS we performed  a   parameter  determination and  a 
detailed abundance  analysis of thirteen potential COROT  main targets.
The precision of our results is very high. Four stars in the preliminary list 
of COROT main targets could not   be analysed with  APASS. The $\delta$  
Scuti  stars  HD 181555   and 
HD 171234 have a  projected  rotational  velocity  
that is too high for our spectroscopic method. For $v$ sin$i$ larger than 150 km/s, 
it becomes very difficult to adjust the continuum correctly and to find
a significant number of well--defined AnU, making APASS unsuitable for 
spectroscopic analysis. On the other hand, the $\beta$ Cep  stars 
HD 170580  and HD 180642 are too different from the Sun
for our differential method. Furthermore, LTE becomes a poor 
assumption for such hot stars.

From the abundance pattern of the analysed stars, we found  no evidence 
of chemical peculiarities.
For   most  of  the  stars, our results  for the fundamental parameters
agree both with the estimates from TEMPLOGG using Str\"omgren photometry and
with results obtained by Bruntt using his software VWA. The  precision of
our results is  high, although we  take the 
influence  of  the  uncertainty  on  the broadening parameters into  account  
 in our computation of the error bars.

We also made an analysis of five COROT main targets, each with 
a moderate rotational velocity, using a classical equivalent--width method.
The results show a clear discrepancy between the two methods. We than 
investigated this  discrepancy and  found that  a classical  equivalent--width 
method using Gaussian or Voigt profile--fitting is not adapted  
to  the analysis  of stars with a projected rotational velocity that is 
significantly different from the solar one. 

For the star   HD 52265,  which  has a massive planet orbiting 
around it and which was analysed by Santos et  al. (2004)  using  a  classical 
equivalent--width method, our  results  agree well with  
the results obtained by  these authors. Nevertheless, comparison  of 
the results obtained after analysing spectra coming   from  the  same  
instrument (CORALIE) seems to  confirm  that an equivalent--width method 
using Gaussian or Voigt profiles tends to give systematic errors   on 
the  parameters and  abundances due  to the sensitivity of the measured  
equivalent--width to the broadening of the  spectrum.  The 
use of rotationally broadened profile could lead to  more reliable results, 
but only in the case of a rather slow rotation. For stars with 
a significantly higher projected  rotational  velocity  than the Sun's, the concept of
isolated lines is lost and only synthetic spectra methods such as APASS 
should be used to spectroscopically determine  the atmosphere
parameters and abundances.  
 
%
%
\begin{acknowledgements}
This research made use of model atmospheres calculated with   ATLAS9 
at the Department of Astronomy of  the University  of  Vienna,  Austria, 
available at http://ams.astro.univie.ac.at/nemo/. We are grateful   to
E. Solano and C. Catala, who are in charge of the Gaudi database. We would like to thank
the authors of the co-addition HARPS spectrum of HD  49933,  B. 
Mosser and C. Catala. We are particularly grateful to C. Catala for  his 
numerous valuable comments and suggestions. This work was  supported 
by the Prodex-ESA Contract 15448/01/NL/Sfe(IC).
\end{acknowledgements}
%
%

%
%

\end{document}